# The modified quasichemical model in the Distinguishable-Pair Approximation for multicomponent solutions

Kun Wang[1*], Patrice Chartrand[2]

*1 State Key Laboratory of Advanced Special Steel & Shanghai Key Laboratory of Advanced Ferrometallurgy & School of Materials Science and Engineering, Shanghai University, Shanghai 200072, China*

*2 Center for Research in Computational Thermochemistry (CRCT), Department of Chemical Engineering, Polytechnique Montréal, Montréal H3C 3A7, Québec, Canada*

Abstract: The Modified Quasichemcial Model in the Distinguishable-Pair Approximation (MQMDPA) for manifold short-range orders in liquids has been successfully extended to multicomponent solutions. The extension is conducted by means of the geometrical interpolation method. Three types of interpolation models, namely Kohler, Toop and Chou, are introduced to initially formulate the pair interaction energies in ternary solutions by employing those in their constituent binary solutions. The pair energies can be expanded in terms of the pair fractions (configuration-dependent) or in terms of the "coordination-equivalent" fractions (composition- dependent). These methods are subsequently extended for use in multicomponent solutions. A general formalism for the combined Kohler-Toop model is employed to permit complete freedom of choice to treat any ternary subsystems with a symmetric or asymmetric model. Meanwhile, a general Chou model is also used to treat all ternary subsystems without any human interference to select symmetric or asymmetric components but only dependent upon the similarity and difference in properties from each two binary solutions with a co-member. Advantages and shortcomings are critically discussed regarding the utilization of different interpolation models.

## 1. Introduction

In the preceding article [1], we proposed the MQMDPA in order to have the capacity to treat manifold short-range orders in liquids within the framework of the quasichemcial approach. The MQMDPA was the further modification to the modified quasichemcial model in the pair approximation (MQMPA) developed by Pelton [2]. The modification was performed based upon the concept of unity and opposites of ordered pairs.

The ordered pairs were initially considered to be distinguishable, emphasizing the aspect of opposites among them. This aspect resulted in the derivation of expression for the configurational entropy contributed from the distinguishable ordered pairs, which is thus responsible for the description of manifold short-range orders in solutions. Interestingly, once the distinguishable ordered pairs are assigned with the same pair interaction energies and coordination numbers, they are automatically transformed to be indistinguishable ordered pairs. The unity of

---

[*]Corresponding author.

E-mail address: wangkun0808@shu.edu.cn

opposites for the ordered pairs is thus conceptualized at this moment. In the aspect of unity among ordered pairs, the MQMDPA has already been reduced to the MQMPA due to their indistinguishable behavior. The MQMDPA can also approach the ideal entropy of mixing if the pair interaction energies tend to be negligible regardless of how to set the coordination numbers. This is the main superiority that the MQMDPA possesses compared to the Associate Solution Model (ASM) [3-5]. As a matter of fact, no interactions between alloying atoms cause unique coordination environments to be formed, which can be viewed as another aspect of unity among ordered pairs. Although the MQMDPA has a lot of merits and is flexible to treat various melt configurations, it has to be extended for use in multicomponent solutions; otherwise its practical applications will be limited.

In the present article, three interpolation methods, namely Kohler [6], Toop [7], and Chou [8], are initially introduced to build the expressions for the pair interaction energies in ternary solutions by using the corresponding parameters optimized from the constituent binary solutions. Later, a generic formalism for the combined Kohler-Toop model proposed by Pelton [9-11] is employed to permit complete freedom of choice to treat any ternary subsystems with a symmetric or asymmetric model. A general Chou model [12] is also used to treat all ternary subsystems without any human interference to arrange components into symmetric or asymmetric groups but only dependent upon the similarity and difference in properties from each two binary solutions with a co-member. The introduced similarity coefficient to characterize the deviation of properties from each two binary solutions with a co-member provides the model to be dynamically shiftable among different interpolation models. These interpolation methods have been successfully applied to the pair energies in terms of the pair fractions or in terms of the coordination-equivalent fractions, which provides the MQMDPA with great flexibility to treat multicomponent alloy solutions or common-ion salt and oxide solutions in various structural configurations. In a third article [13], the MQMDPA will be further extended to treat multicomponent reciprocal solutions within the two-sublattice quadruplet approximation [14].

## 2. The Model

In the preceding article [1], we proposed the MQMDPA formalism to directly treat binary solutions possessing multiple compositions of Maximum Short-Range Ordering (MSRO). The new formalism can overcome many uncertain problems resulting from the combinatorial model [15-16] of MQMPA and MSM in thermodynamically describing such solutions. These problems included but not limited to groundless definitions of lattice stabilities for multiple state species, inexplicit choice of interpolation model for pseudo multicomponent solutions, and many more model parameters required. In view of these merits, the present paper considers extending the MQMDPA to describe multicomponent solutions. The expansion formalism is expressed as,

$$G = \sum n_m g_m^0 - T\Delta S^{config} + \sum_{\langle k \rangle} \sum_{m\ >\ n} \sum n_{mn}^{\langle k \rangle} \Delta g_{mn}^{\langle k \rangle} \qquad (1)$$

where $n_m$ and $g_m^0$ are the number of moles of and the molar Gibbs energy of pure component *m*, T refers to the temperature in kelvin, $\Delta S^{config}$ stands for the configurational entropy of mixing, and $n_{mn}^{\langle k \rangle}$ and $\Delta g_{mn}^{\langle k \rangle}$ represent the number of moles of and the molar Gibbs energy change for the formation of the *m-n* pair with index *k* denoting the distinguishable types of ordered pairs. If there are two types of *m-n* pairs in solution, their pair exchange reactions and the resultant Gibbs energy change for pair reactions are shown as,

$$(m - m) + (n - n) = 2(m - n) \quad \Delta g_{mn}^{(1)} \qquad (m - m) + (n - n) = 2(m = n) \quad \Delta g_{mn}^{(2)} \quad \ldots \quad (2)$$

where ($m$ and $n$) are equal to 1, 2 … N, and ($m$-$m$, $n$-$n$, and $m$-$n$) represent first-nearest-neighbor (FNN) pairs. $\Delta S^{config}$ is an approximate expression for the configurational entropy of mixing as the following equation shows,

$$-\frac{\Delta S^{config}}{R} = \sum n_m ln(X_m) + \sum n_{mm} \ln\left(\frac{X_{mm}}{Y_m^2}\right) + \sum_{m > n} \sum_{\langle k \rangle} n_{mn}^{\langle k \rangle} \ln\left(\frac{N_{mn} X_{mn}^{\langle k \rangle}}{2 Y_m Y_n}\right) \tag{3}$$

where it consists of three parts: the first term denoting the entropy from ideally mixing atoms or pairs, the second one regarding mixing the homogenous pairs ($m$-$m$, $n$-$n$…), and the last term considering mixing the $N_{mn}$ types of heterogeneous $m$-$n$ pairs. If there are three types of $m$-$n$ pairs in solution, $N_{mn}$ is thus equivalent to three and the index $k$ can be assigned with 1, 2 or 3. The mole fractions of atom (or molecule) $m$, homogenous pair $m$-$m$ and heterogeneous pair $m$-$n$ are respectively expressed as equations (4-6),

$$X_m = n_m / \sum n_i \tag{4}$$

$$X_{mm} = n_{mm} / (\sum n_{ii} + \sum_{i \neq j} n_{ij}^{\langle k \rangle}) \tag{5}$$

$$X_{mn}^{\langle k \rangle} = n_{mn}^{\langle k \rangle} / (\sum_i n_{ii} + \sum_{i \neq j} n_{ij}^{\langle k \rangle}) \tag{6}$$

The "coordination-equivalent" fraction is defined as equation (7),

$$Y_m = \frac{Z_m n_m}{\sum Z_i n_i} = \frac{Z_m X_m}{\sum Z_i X_i} \tag{7}$$

where $Z_m$ stands for the overall coordination number of atom (or molecule) $m$. The number of moles of atom (or molecule) $m$ could be related to those of the corresponding pairs using the coordination number as equation (8) shows,

$$Z_m n_m = 2 n_{mm} + \sum_{\langle k \rangle} \sum_{m \neq n} n_{mn}^{\langle k \rangle} \tag{8}$$

By taking equation (8) into equation (7), the mass relation between the "coordination-equivalent" fraction and the relative pair fractions is established as,

$$Y_m = X_{mm} + \sum_{\langle k \rangle} \sum_{n \neq m} \frac{X_{mn}^{\langle k \rangle}}{2} \tag{9}$$

where $X_{mn}^{\langle k \rangle}$ represents the pair fraction of the $m$-$n$ pair in the $k$ type. The ratio of coordination numbers $Z_m/Z_n$ determines the chemical composition where the MSRO occurs in the binary $m$-$n$ solution. In the ternary $l$-$m$-$n$ solution, if the coordination numbers are all constant, there will be no free choice of the chemical composition of the MSRO in the third $l$-$n$ solution since $Z_l/Z_n$ is predefined when the $l$-$m$ and $m$-$n$ solutions have the MSROs at the chemical compositions of $Z_l/Z_m$ and $Z_m/Z_n$. To overcome this issue, the overall coordination numbers are defined as,

$$\frac{1}{Z_m} = \frac{1}{2n_{mm} + \sum_{\langle k \rangle} \sum_{n \neq m} n_{mn}^{\langle k \rangle}} \left( \frac{2n_{mm}}{Z_{mm}^m} + \sum_{\langle k \rangle} \sum_{n \neq m} \frac{n_{mn}^{\langle k \rangle}}{Z_{mn}^{\langle k \rangle m}} \right) \tag{10}$$

$$\frac{1}{Z_n} = \frac{1}{2n_{nn} + \sum_{\langle k \rangle} \sum_{m \neq n} n_{mn}^{\langle k \rangle}} \left( \frac{2n_{nn}}{Z_{nn}^n} + \sum_{\langle k \rangle} \sum_{m \neq n} \frac{n_{mn}^{\langle k \rangle}}{Z_{mn}^{\langle k \rangle n}} \right) \tag{11}$$

where $Z_{mm}^m$ and $Z_{mn}^{\langle k \rangle m}$ are, respectively, the values of $Z_m$ when all nearest neighbors of an $m$ are also $m$ and when all nearest neighbors of the $m$ are $n$ forming all $m$-$n$ pairs in the $k$ type. The term $Z_{mm}^m$ is constant for each pure component $m$ and is the same for all solutions containing $m$. The chemical composition of MSRO in each binary subsystem is determined by the ratio $Z_{mn}^{\langle k \rangle m}/Z_{mn}^{\langle k \rangle n}$. Taking equations (10-11) into equation (8) reads,

$$n_m = \frac{2n_{mm}}{Z_{mm}^m} + \sum_{n \neq m} \sum_k \frac{n_{mn}^{\langle k \rangle}}{Z_{mn}^{\langle k \rangle m}} \tag{12}$$

which clearly reflects the true mass balance between component $m$ and the $m$ related pairs. According to the mass balance relations, the Gibbs energies of pairs ($g_{mm}$ and $g_{mn}$) are defined as,

$$g_{mm} = g_{mm}^0 = \frac{2g_m^0}{Z_{mm}^m} \tag{13}$$

$$g_{mn}^{\langle k \rangle} = \frac{g_m^0}{Z_{mn}^{\langle k \rangle m}} + \frac{g_n^0}{Z_{mn}^{\langle k \rangle n}} + \Delta g_{mn}^{\langle k \rangle} \tag{14}$$

where the superscript "0" denotes the standard Gibbs energy which is composition independent. It is obvious that the Gibbs energy of the homogenous pair $m$-$m$ is the same as its standard Gibbs energy. Substitution of equations (13-14) into equation (1) then gives,

$$G = \sum n_{mm} g_{mm}^0 + \sum_{m > n} \sum_k n_{mn}^{\langle k \rangle} g_{mn}^{\langle k \rangle} - T\Delta S^{config} + \sum_m \sum_{> n} \sum_{\langle k \rangle} n_{mn}^{\langle k \rangle} \Delta g_{mn}^{\langle k \rangle} \tag{15}$$

where $\Delta g_{mn}^{\langle k \rangle}$ could be expressed as an empirical polynomial either in the "coordination-equivalent" fractions or in the pair fractions. In the "coordination-equivalent" fractions, $\Delta g_{mn}^{\langle k \rangle}$ has the formalism as,

$$\Delta g_{mn}^{\langle k \rangle} = \Delta g_{mn}^{0\langle k \rangle} + \sum_{(i+j) \geq 1} q_{mn}^{ij\langle k \rangle} Y_m^i Y_n^j \tag{16}$$

while in the pair fractions, $\Delta g_{mn}^{\langle k \rangle}$ has the expression as,

$$\Delta g_{mn}^{\langle k \rangle} = \Delta g_{mn}^{0\langle k \rangle} + \sum_{(i+j) \geq 1} q_{mn}^{ij\langle k \rangle} X_{mm}^i Y_{nn}^j \tag{17}$$

where $\Delta g_{mn}^{0\langle k \rangle}$ and $q_{mn}^{ij\langle k \rangle}$ are empirical binary coefficients, which may be functions of temperature and pressure. Equation (17) is termed by Blander [17] as the "configuration-dependent" expression. It has already demonstrated that equation (17) is more convenient to be used to fit the experimental data with fewer coefficients than equation (16). However, equations (16-17) only apply in the binary subsystems. It is now required to write expressions for $\Delta g_{mn}^{\langle k \rangle}$ for compositions within the multicomponent system for use in equation (15). In the Calphad community, this conversion process is called as the geometrical interpolation method, as discussed in detail in the following section.

## 3. The interpolation schemes

In this section, several geometrical interpolation methods will be used to convert $\Delta g_{mn}^{\langle k \rangle}$ from binary expression to multicomponent formalism in terms of composition. At the beginning, we will demonstrate how to transform binary $\Delta g_{mn}^{\langle k \rangle}$ expressed in "coordination-equivalent" fractions into ternary and multicomponent ones by using the Kohler model [6], Toop model [7] and Chou model [8], respectively. Subsequently, binary "configuration-dependent" $\Delta g_{mn}^{\langle k \rangle}$ will be expanded into ternary and multicomponent ones following the similar approach [9-12]. Last but not the least, pros and cons among the interpolation models will be fully discussed and the possible code implementation is suggested.

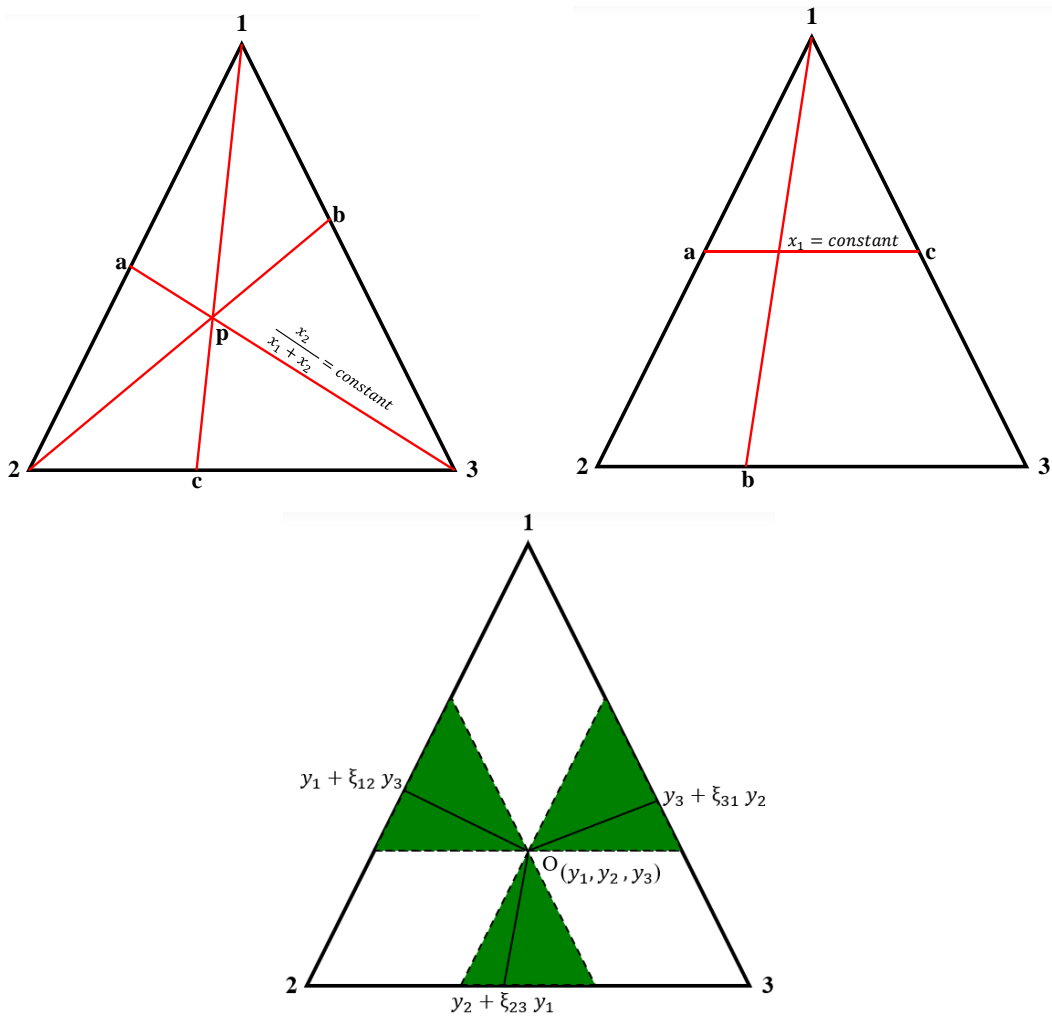

Fig.1 Some geometrical interpolation methods: (a) symmetrical Kohler model; (b) asymmetrical Toop model; (c) integration model

## 3.1 Kohler model

If $\Delta g_{12}^{\langle k \rangle}$ in the binary 1-2 subsystem has been expressed as a polynomial in the "coordination-equivalent" fractions $Y_1$ and $Y_2$ by equation (16), then the corresponding expression in the ternary 1-2-3 system is given by,

$$\Delta g_{12}^{\langle k \rangle} = (\Delta g_{12}^{\langle k \rangle 0} + \sum_{i+j \geq 1} q_{12}^{\langle k \rangle ij} \left(\frac{Y_1}{Y_1 + Y_2}\right)^i \left(\frac{Y_2}{Y_1 + Y_2}\right)^j ) + \sum_{\substack{m \geq 1 \\ i \geq 0 \\ j \geq 0}} q_{12(3)}^{ijm} \left(\frac{Y_1}{Y_1 + Y_2}\right)^i \left(\frac{Y_2}{Y_1 + Y_2}\right)^j Y_3^m \tag{18}$$

where the first term on the right-hand side of equation (18) is constant along the line *3-a* in figure 1(a) and is equal to $\Delta g_{12}^{\langle k \rangle}$ in the binary at point *a* with the variables limiting to $Y_1+Y_2=1$. This means the binary 1-2 pair interaction energy is constant at a constant $Y_1/Y_2$ ratio. The second summation in equation (18) comprises of "ternary terms" that are all ZERO in the 1-2 binary, and which give the effect of the presence of component 3 upon the energy $\Delta g_{12}^{\langle k \rangle}$ of the *k*-type 1-2 pair interactions. The empirical ternary coefficients $q_{12(3)}^{ijm}$ are found by optimization of experimental ternary data.

If $\Delta g_{12}^{\langle k \rangle}$ in the binary 1-2 subsystem has been formulated by a polynomial in terms of the pair interactions as equation (17) displays, then the following equation for $\Delta g_{12}^{\langle k \rangle}$ is proposed,

$$\Delta g_{12}^{\langle k \rangle} = \Delta g_{12}^{\langle k \rangle 0} + \sum_{i+j \geq 1} q_{12}^{\langle k \rangle ij} \left(\frac{X_{11}}{X_{11} + X_{22} + \sum_{\langle k \rangle} X_{12}^{\langle k \rangle}}\right)^i \left(\frac{X_{22}}{X_{11} + X_{22} + \sum_{\langle k \rangle} X_{12}^{\langle k \rangle}}\right)^j$$

$$+ \sum_{\substack{m \geq 1 \\ i \geq 0 \\ j \geq 0}} q_{12(3)}^{ijm} \left(\frac{X_{11}}{X_{11} + X_{22} + \sum_{\langle k \rangle} X_{12}^{\langle k \rangle}}\right)^i \left(\frac{X_{22}}{X_{11} + X_{22} + \sum_{\langle k \rangle} X_{12}^{\langle k \rangle}}\right)^j Y_3^m \tag{19}$$

where $\Delta g_{12}^{\langle k \rangle 0}$ and $q_{12}^{\langle k \rangle ij}$ are optimized by using experimental data from binary 1-2 system, and $q_{12(3)}^{ijm}$ is done from ternary 1-2-3 system. The formalism is expressed by considering the following facts. As $\Delta g_{12}^{\langle k \rangle}$, $\Delta g_{23}^{\langle k \rangle}$ and $\Delta g_{13}^{\langle k \rangle}$ become small, the solution approaches ideality, and $X_{11} \to Y_1^2$ $X_{22} \to Y_2^2$ and $\sum_{\langle k \rangle} X_{12}^{\langle k \rangle} \to 2Y_1 Y_2$. In this case, $X_{11}/(X_{11} + X_{22} + \sum_{\langle k \rangle} X_{12}^{\langle k \rangle}) \to (Y_1/(Y_1 + Y_2))^2$, and equation (19) approaches equation (18), which, in the limit, becomes the well-known Kohler [6] equation for symmetrical ternary systems.

Similar equations give $\Delta g_{23}^{\langle k \rangle}$ and $\Delta g_{31}^{\langle k \rangle}$, with the binary terms equal to their values at point b and c, respectively, in Figure 1(a) and with the ternary coefficients $q_{23(1)}^{ijm}$ and $q_{31(2)}^{ijm}$, which give the effect of the presence of component 1 upon the pair energy $\Delta g_{23}^{\langle k \rangle}$ and of component 2 upon $\Delta g_{31}^{\langle k \rangle}$, respectively. This model is "symmetric" since the three components are treated in the same fashion. From equation (9), it can be seen that the factor $Y_3$ in the ternary terms in equation (19) is equal to $(X_{33} + \sum_k X_{31}^{\langle k \rangle}/2 + \sum_{k'} X_{23}^{\langle k' \rangle}/2)$. In principle, the effect of these three terms upon $\Delta g_{12}^{\langle k \rangle}$ could easily be represented by three independent ternary coefficients. However, this additional complexity is probably not required.

## 3.2 Toop model

For certain systems, there exists one component which is chemically different from the other two, such as $SiO_2$-$CaO$-$MgO$, S-Fe-Cu, Na-Au-Ag, etc., it is more appropriate to use the asymmetric model illustrated in Figure 1(b), where component 1 is the asymmetric component. In this case, if binary $\Delta g_{12}^{\langle k \rangle}$ has been expressed as polynomial in "coordination-equivalent" fractions, the ternary $\Delta g_{12}^{\langle k \rangle}$ can be given by,

$$\Delta g_{12}^{\langle k \rangle} = (\Delta g_{12}^{\langle k \rangle 0} + \sum_{i+j \geq 1} q_{12}^{ij} Y_1^i (1 - Y_1)^j)) + \sum_{\substack{m \geq 1 \\ i \geq 0 \\ j \geq 0}} q_{12(3)}^{ijm} Y_1^i (1 - Y_1)^j (\frac{Y_3}{Y_2 + Y_3})^m \quad (20)$$

where the binary terms are constant along the line *ac* and equal to their values at point *a* in figure 1(b). A similar expression is written for $\Delta g_{31}^{\langle k \rangle}$, while $\Delta g_{23}^{\langle k \rangle}$ is given by an expression similar to equation (18). If binary $\Delta g_{12}^{\langle k \rangle}$ has been expressed as polynomial in pair fractions, the following equation is proposed,

$$\Delta g_{12}^{\langle k \rangle} = \Delta g_{12}^{\langle k \rangle 0} + \sum_{i+j \geq 1} q_{12}^{\langle k \rangle ij} X_{11}^{i} (X_{22} + X_{33} + \sum_{\langle k \rangle} X_{23}^{\langle k \rangle})^j + \sum_{\substack{m \geq 1 \\ i \geq 0 \\ j \geq 0}} q_{12(3)}^{ijm} X_{11}^{i} (X_{22} + X_{33} + \sum_{\langle k \rangle} X_{23}^{\langle k \rangle})^j (\frac{Y_3}{Y_2 + Y_3})^m \quad (21)$$

In the limit of ideality, equation (21) reduces to equation (20), which, in the limit, becomes the well-known Kohler-Toop equation [7] for asymmetrical ternary systems. It has been shown that, for systems with large composition-dependent deviations from ideality, the choice of a symmetric or asymmetric model can often give very different results. An incorrect choice can even give rise to spurious miscibility gaps.

### 3.3 Chou model

It is clearly seen that there is much difference in the fashion of the geometrical interpolations using the symmetrical Kohler model and the asymmetrical Toop model. Hence, different selections of models and asymmetrical components in a ternary system would lead to different results of the ternary Gibbs free energy of mixing, in particular having the subsystems of which Gibbs free energies of mixing are strongly composition dependent [9]. In order to overcome this uncertainty, we usually select the interpolation model and arrange the components depending on certain values, such as the chemical properties of components, the location of elements in a periodic table [16], or the valence of compounds [18-19]. In the words of Chou [8], the Kohler-Toop model needs human interference to decide how to define symmetric and asymmetric components. Chou [8] then proposed the integration model, which was expected to avoid the human interference, and tried to define binary composition for geometrical interpolation by using similarity coefficients. The similarity coefficients are defined as,

$$\xi_{12} = \frac{\eta_\mathrm{I}}{\eta_\mathrm{I} + \eta_\mathrm{II}} \qquad \xi_{23} = \frac{\eta_\mathrm{II}}{\eta_\mathrm{II} + \eta_\mathrm{III}} \qquad \xi_{31} = \frac{\eta_\mathrm{III}}{\eta_\mathrm{III} + \eta_\mathrm{I}} \quad (22)$$

where $\eta_\mathrm{I}$, $\eta_\mathrm{II}$ and $\eta_\mathrm{III}$ are the so-called "deviation sum of squares" and given as,

$$\eta_\mathrm{I} = \int_0^1 (\Delta G_{12} - \Delta G_{13})^2 dn_1 \qquad \eta_\mathrm{II} = \int_0^1 (\Delta G_{21} - \Delta G_{23})^2 dn_2 \qquad \eta_\mathrm{III} = \int_0^1 (\Delta G_{31} - \Delta G_{32})^2 dn_3 \quad (23)$$

where $\Delta G_{ij}$ should be expressed as the quasichemical formalism. Obviously, if the component "3" is similar to the component "2" thermodynamically, the value of $\eta_\mathrm{I}$ should approach zero, compelling $\xi_{12}$ to be zero as well. If the component "3" is similar to the component "1" thermodynamically, the value of $\eta_\mathrm{II}$ should approach zero, forcing $\xi_{12}$ to be unity. Therefore, the $\xi_{12}$ value ranges from 0 to 1, which can judge if the third component is more similar to the component one or two. The similar situation could be found for $\xi_{23}$ and $\xi_{31}$. The three similarity coefficients for three binaries are correlated as the following expression,

$$(1 - \xi_{12})(1 - \xi_{31})(1 - \xi_{23}) = \xi_{12} \xi_{31} \xi_{23} \quad (24)$$

The above equation provides the mechanism to calculate the third similarity coefficient from the other two known coefficients. Based upon the above definitions, the integration model will select the following binary compositions (coordination-equivalent fraction),

$$Y_{1(12)} = Y_1 + \xi_{12}Y_3 \qquad Y_{2(23)} = Y_2 + \xi_{23}Y_1 \qquad Y_{3(31)} = Y_3 + \xi_{31}Y_2 \qquad (25)$$

which is schematically shown in Figure 1c. Since all the ξ values vary from 0 to 1, the lines with the two ends connecting the ternary composition and its interpolated binary composition sweep over all the green zones where the integration model is ergodic over all the Kohler, Muggianu and the Toop models. For instance, assuming $\xi_{12}$ to be equal to 0.5, the integration model is then shifted to the Muggianu model for selecting the binary composition.

Suppose $\Delta g_{12}^{\langle k \rangle}$ has been expressed as equation (16), the corresponding formalism in the ternary 1-2-3 system can be given as,

$$\Delta g_{12}^{\langle k \rangle} = \Delta g_{12}^{\langle k \rangle 0} + \sum_{i+j \geq 1} q_{12}^{\langle k \rangle ij}(Y_1 + \xi_{12}Y_3)^i(Y_2 + \xi_{12}Y_3)^j + \sum_{i+j \geq 1} q_{123}^{\langle k \rangle ijm}(Y_1 + \xi_{12}Y_3)^i(Y_2 + \xi_{12}Y_3)^j Y_3^m \qquad (26)$$

where $i$ or $j$ are assigned with even values, there will be $Y_1^2$, $Y_2^2$ and $2Y_1Y_2$ approaching $X_{11}$, $X_{22}$ and $\sum_{\langle k \rangle} X_{12}^{\langle k \rangle}$ in the limit of ideality, respectively. This can result in the following expression as,

$$\Delta g_{12}^{\langle k \rangle} = \Delta g_{12}^{\langle k \rangle 0} + \sum_{i+j \geq 1} q_{12}^{\langle k \rangle ij}[X_{11} + \xi_{12}^2 X_{33} + \xi_{12} \sum_{\langle k \rangle} X_{13}^{\langle k \rangle}]^i [X_{22} + (1-\xi_{12})^2 X_{33} + (1-\xi_{12}) \sum_{\langle k \rangle} X_{23}^{\langle k \rangle}]^j +$$
$$\sum_{\substack{m \geq 1 \\ i \geq 0 \\ j \geq 0}} q_{12(3)}^{ijm}[X_{11} + \xi_{12}^2 X_{33} + \xi_{12} \sum_{\langle k \rangle} X_{13}^{\langle k \rangle}]^i [X_{22} + (1-\xi_{12})^2 X_{33} + (1-\xi_{12}) \sum_{\langle k \rangle} X_{23}^{\langle k \rangle}]^j Y_3^m \qquad (27)$$

which is actually the 1-2 pair interaction energy expanded by the pair fractions in the 1-2-3 ternary system.

The similarity coefficient $\xi_{12}$ has to be pre-calculated using the integral equation (23) along with equation (22). Since $\Delta G_{ij}$ is now written in the quasichemical formalism, it is infeasible to exactly solve the integral due to the existing internal variables (pair fractions). In binary systems, the pair fractions are initially obtained by minimizing the Gibbs energy at fixed composition, temperature and pressure, and then they are taken into the Gibbs energy expression. However, it is not sure if the pair fractions could be taken into equation (23) to obtain the minimal $\eta$ values. This uncertainty could be well clarified by considering the following expressions. Let $F(n_1, n_{12}, n_{13})$ be the integrand as,

$$F(n_1, n_{12}, n_{13}) = [\Delta G_{12}(n_1, n_{12}) - \Delta G_{13}(n_1, n_{13})]^2 \qquad (28)$$

and $\eta$ will have the expression as,

$$\eta_I = \int_0^1 F(n_1, n_{12}, n_{13}) dn_1 \qquad (29)$$

Minimizing η at fixed composition, the following equation yields,

$$\frac{\partial \eta_I}{\partial n_{12}} = \int_{0=f(n_{12},n_{13})}^{1=g(n_{12},n_{13})} \frac{\partial F(n_1, n_{12}, n_{13})}{\partial n_{12}} dn_1 + \left[\frac{\partial g}{\partial n_{12}} - \frac{\partial f}{\partial n_{12}}\right] F(n_1, n_{12}, n_{13}) = 0 \qquad (30)$$

$$\frac{\partial \eta_I}{\partial n_{13}} = \int_{0=f(n_{12},n_{13})}^{1=g(n_{12},n_{13})} \frac{\partial F(n_1, n_{12}, n_{13})}{\partial n_{13}} dn_1 + \left[\frac{\partial g}{\partial n_{13}} - \frac{\partial f}{\partial n_{13}}\right] F(n_1, n_{12}, n_{13}) = 0 \qquad (31)$$

where the second terms on the right-hand side are always zero. This indicates that the following equations must result,

$$\frac{\partial F(n_1,n_{12},n_{13})}{\partial n_{12}} = 2[\Delta G_{12}(n_1,n_{12}) - \Delta G_{13}(n_1,n_{13})]\left[\frac{\partial \Delta G_{12}(n_{12},n_1)}{\partial n_{12}}\right] = 0 \tag{32}$$

$$\frac{\partial F(n_1,n_{12},n_{13})}{\partial n_{13}} = 2[\Delta G_{12}(n_1,n_{12}) - \Delta G_{13}(n_1,n_{13})]\left[\frac{\partial \Delta G_{12}(n_{12},n_1)}{\partial n_{13}}\right] = 0 \tag{33}$$

since $n_1$ is an arbitrary value ranging from 0 to 1 in equations (30-31). In general, the deviation between $\Delta G_{12}$ and $\Delta G_{13}$ should not always disappear, and the following expressions are thus generated,

$$\frac{\partial \Delta G_{12}(n_{12},n_1)}{\partial n_{12}} = 0 \qquad \frac{\partial \Delta G_{12}(n_{12},n_1)}{\partial n_{13}} = 0 \tag{34}$$

It is evident that equation (34) is the right minimization condition to obtain the equilibrium pair fractions for the binary 1-2 and 1-3 systems. This indicates that the similarity coefficients could be calculated by just taking the equilibrium pair fractions obtained from each binary subsystem into the integral equations. The integral equations, such as $\eta_I$, can be approximated by its closely related sum,

$$\eta_I = \int_0^1 (\Delta G_{12} - \Delta G_{13})^2 dn_1 \cong \sum_{i=1}^{1/\Delta h - 1} \Delta h [\Delta G_{12}(i\Delta h, \dot{X}_{12}) - \Delta G_{13}(i\Delta h, \dot{X}_{13})]^2 \tag{35}$$

where $\dot{X}_{12}$ and $\dot{X}_{13}$ are the equilibrium fractions of pairs 1-2 and 1-3 in the binary 1-2 and 1-3 subsystems at compositions $i\Delta h$, respectively. After numerous tests, it is found that $\Delta h$ equal to 0.1 could yield good approximations. The similar approximation could be used for calculating $\eta_{II}$ and $\eta_{III}$.

### 3.4 Multicomponent solutions

The MQMDPA can also be extended to multicomponent solutions using the method proposed by Pelton [9-11]. The interpolation method combines the Kohler and Toop models, showing complete flexibility to treat any ternary subsystem as symmetric or asymmetric. The selected binary composition is defined as,

$$Y_{a(ab)} = Y_a + \sum_c Y_c \tag{36}$$

where the summation is over all *c* components in asymmetric *a-b-c* ternary subsystem in which *b* is the asymmetric component. Taking equation (36) into equation (16), the *a-b* pair interaction energy in multicomponent system reads,

$$\Delta g_{ab}^{\langle k \rangle} = \Delta g_{ab}^{\langle k \rangle 0} + \sum_{i+j\geq 1} q_{ab}^{ij} \left(\frac{Y_{a(ab)}}{Y_{a(ab)}+Y_{b(ab)}}\right)^i \left(\frac{Y_{b(ab)}}{Y_{a(ab)}+Y_{b(ab)}}\right)^j + \sum_{\substack{o\geq 1 \\ i\geq 0 \\ j\geq 0}} \left(\frac{Y_{a(ab)}}{Y_{a(ab)}+Y_{b(ab)}}\right)^i \left(\frac{Y_{b(ab)}}{Y_{a(ab)}+Y_{b(ab)}}\right)^j \left(\sum_l q_{ab(l)}^{ijo} Y_l (1 - Y_{a(ab)} - Y_{b(ab)})^{o-1} + \sum_e q_{ab(e)}^{ijo}\left(\frac{Y_e}{Y_{b(ab)}}\right)(1 - \frac{Y_b}{Y_{b(ab)}})^{o-1} + \sum_f q_{ab(f)}^{ijo}\left(\frac{Y_f}{Y_{a(ab)}}\right)\left(1 - \frac{Y_a}{Y_{a(ab)}}\right)^{o-1}\right) \tag{37}$$

where in the limit of ideality, $Y_a^2, Y_c^2, 2Y_aY_c$ and $2Y_cY_{c'}$ will get close to $X_{aa}, X_{cc}, X_{ac}$ and $X_{cc'}$. This can result equation (37) in the following form,

$$\Delta g_{ab}^{\langle k \rangle} = \Delta g_{ab}^{\langle k \rangle 0} + \sum_{i+j\geq 1} q_{ab}^{ij} \chi_{ab}^i \chi_{ba}^j + \sum_{\substack{o\geq 1 \\ i\geq 0 \\ j\geq 0}} \chi_{ab}^i \chi_{ba}^j \left(\sum_l q_{ab(l)}^{ijo} Y_l (1 - Y_{a(ab)} - Y_{b(ab)})^{o-1} + \sum_e q_{ab(e)}^{ijo}\left(\frac{Y_e}{Y_{b(ab)}}\right)(1 - \frac{Y_b}{Y_{b(ab)}})^{o-1} + \sum_f q_{ab(f)}^{ijo}\left(\frac{Y_f}{Y_{a(ab)}}\right)(1 - \frac{Y_i}{Y_{a(ab)}})^{o-1}\right) \tag{38}$$

In equations (37-38), the summations of ternary terms are over (1) all *l* components in 1-2-*l* ternary subsystems, which are either symmetric or in which l is the asymmetric component; (2) all *e* components in 1-2-*e* subsystems, in which 1 is the asymmetric component; (3) all $f$ components in 1-2-$f$ subsystems, in which 2 is the asymmetric component. $\chi_{ab}$ and $\chi_{ba}$ in equation (38) are with the following expression,

$$\chi_{ab} = \frac{X_{aa}+\sum_c\sum_{\langle k \rangle}(X_{cc}+X_{ac}^{\langle k \rangle})+\sum_{c>c'}\sum_{\langle k \rangle}X_{cc'}^{\langle k \rangle}}{X_{aa}+\sum_c\sum_{\langle k \rangle}(X_{cc}+X_{ac}^{\langle k \rangle})+\sum_{c>c'}\sum_{\langle k \rangle}X_{cc'}^{\langle k \rangle}+X_{bb}+\sum_d\sum_{\langle k \rangle}(X_{dd}+X_{bd}^{\langle k \rangle})+\sum_{d>d'}\sum_{\langle k \rangle}X_{dd'}^{\langle k \rangle}} \tag{39}$$

$$\chi_{ba} = \frac{X_{bb}+\sum_d\sum_{\langle k \rangle}(X_{dd}+X_{bd}^{\langle k \rangle})+\sum_{d>d'}\sum_{\langle k \rangle}X_{dd'}^{\langle k \rangle}}{X_{aa}+\sum_c\sum_{\langle k \rangle}(X_{cc}+X_{ac}^{\langle k \rangle})+\sum_{c>c'}\sum_{\langle k \rangle}X_{cc'}^{\langle k \rangle}+X_{bb}+\sum_d\sum_{\langle k \rangle}(X_{dd}+X_{bd}^{\langle k \rangle})+\sum_{d>d'}\sum_{\langle k \rangle}X_{dd'}^{\langle k \rangle}} \tag{40}$$

It should be noted here equations (37-38) could be reduced either to the symmetrical equation (20) or asymmetrical equation (21) in any ternary subsystems.

If the Chou model [12] is used to perform the composition interpolation, the binary composition is then selected as,

$$Y_{a(ab)} = Y_a + \sum_{\substack{l=1 \\ l \neq a,b}}^m Y_l \xi_{a(ab)}^{(l)} \tag{41}$$

where the similarity coefficient $\xi_{a(ab)}^{(l)}$ can be calculated as,

$$\xi_{a(ab)}^{(l)} = \frac{\int_0^1 (\Delta G_{ab} - \Delta G_{al})^2 dY_a}{\int_0^1 (\Delta G_{ab} - \Delta G_{al})^2 dY_a + \int_0^1 (\Delta G_{ba} - \Delta G_{bl})^2 dY_b} \tag{42}$$

The integral can still be approximated by using equation (35) for all the binary subsystems. Taking equation (41) into equation (16), the $a$-$b$ pair interaction energy in multicomponent system is with the form as,

$$\Delta g_{ab}^{\langle k \rangle} = \Delta g_{ab}^{\langle k \rangle 0} + \sum_{i+j \geq 1} q_{ab}{}^{ij} Y_{a(ab)}{}^i Y_{b(ab)}{}^j + \sum_{\substack{o \geq 1 \\ i \geq 0 \\ j \geq 0}} \sum_{l \neq a,b} Y_{a(ab)}{}^i Y_{b(ab)}{}^j Y_l^o \, q_{ij(l)}^{ijo} \tag{43}$$

Analogously, the above equation expressed as the "coordination-equivalent" fraction could be transformed to the following equation with the pair fractions,

$$\Delta g_{ab}^{\langle k \rangle} = \Delta g_{ab}^{\langle k \rangle 0} + \sum_{i+j \geq 1} q_{ab}{}^{ij} \chi_{ab}{}^i \chi_{ba}{}^j + \sum_{\substack{o \geq 1 \\ m \geq 0 \\ n \geq 0}} \sum_{l \neq a,b} \chi_{ab}{}^i \chi_{ba}{}^j Y_l^o \, q_{ij(l)}^{mno} \tag{44}$$

where $\chi_{ab}$ and $\chi_{ba}$ are given as,

$$\chi_{ab} = X_{aa} + \sum_l X_{ll} \xi_{a(ab)}^{(l)}{}^2 + \sum_l \sum_{\langle k \rangle} X_{al}^{\langle k \rangle} \xi_{a(ab)}^{(l)} + \sum_{l>l'} \sum_{\langle k \rangle} \xi_{a(ab)}^{(l)} \xi_{a(ab)}^{(l')} X_{ll'}^{\langle k \rangle} \tag{45}$$

$$\chi_{ba} = X_{bb} + \sum_l X_{ll} \xi_{b(ab)}^{(l)}{}^2 \sum_l \sum_{\langle k \rangle} X_{bl}^{\langle k \rangle} \xi_{b(ab)}^{(l)} + \sum_{l>l'} \sum_{\langle k \rangle} \xi_{b(ab)}^{(l)} \xi_{b(ab)}^{(l')} X_{ll'}^{\langle k \rangle} \tag{46}$$

To date, the three interpolation methods have been introduced to expand the pair interaction energies from binary to multicomponent systems both in the "coordination-equivalent" fractions and the pair fractions. The Kohler and Toop models require human interference to arrange components in symmetrical and asymmetrical groups by experience. The interpolation fashion will not be changed once the arrangement is finished. The Chou model introduces similarity coefficients to depict the degree of similarity between components. The similarity coefficients are also dependent upon temperature. They can thus provide dynamic adjustments among different interpolation methods in terms of systems and components themselves. However, the similarity coefficients have to be calculated by integrals in advance using Gibbs energy of mixing from each of all the subsystems at changeable temperatures, which may cause computing algorithm in low execution efficiency, especially in the multicomponent system with large amount of components. On the contrary, the Kohler-Toop model is an analytical expression which can be directly implemented into software without any precalculations of similarity coefficients. The computing algorithm should be in high execution efficiency, and can thus perform fast calculations of thermodynamic properties and phase equilibria in multicomponent solutions.

## 4. Case applications

In this section, thermodynamic properties of a hypothetical ternary solution are calculated using the MQMDPA interpolated by the Kohler, Toop and integration methods, respectively. The calculated results are later compared to show their differences with the three interpolation methods. Assuming the A-B-C ternary system, the subsystems A-B and A-C contains two compositions of MSRO while the subsystem B-C only has one composition of MSRO. The coordination numbers are used as,

$$Z_{AA}^{A} = Z_{BB}^{B} = Z_{CC}^{C} = Z_{AB}^{\langle 1 \rangle A} = Z_{AC}^{\langle 1 \rangle A} = 2Z_{AB}^{\langle 1 \rangle B} = 2Z_{AC}^{\langle 1 \rangle C} = 2Z_{AB}^{\langle 2 \rangle A} = 2Z_{AC}^{\langle 2 \rangle A} = Z_{AB}^{\langle 2 \rangle B} = Z_{AC}^{\langle 2 \rangle C} = Z_{BC}^{\langle 1 \rangle B} = Z_{BC}^{\langle 1 \rangle C} = 4 \quad (47)$$

where the indices $\langle 1 \rangle$ and $\langle 2 \rangle$ refer to the first and second MSRO in the binary solutions.

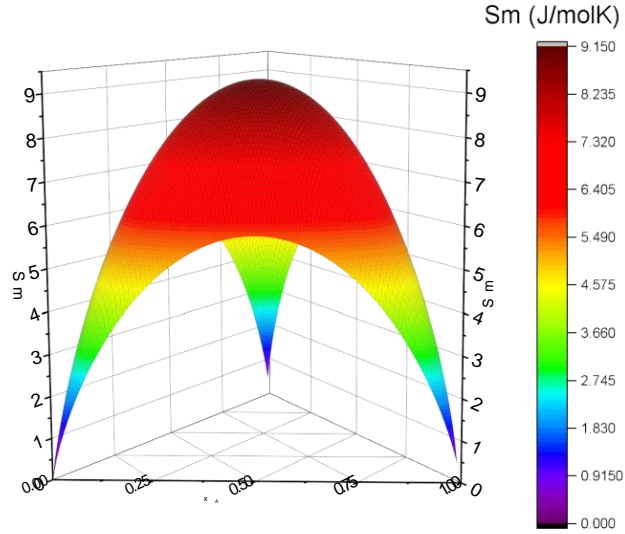

Fig.2 The ideal entropy of mixing over the whole composition space in the ternary A-B-C solution

Suppose all the pair energies are equal to zero. Fig.2 shows the calculated entropy of mixing in the ternary A-B-C solution. It can be seen that the ideal entropy of mixing is achieved, indicating that the configurational entropy is reduced to the one in the Bragg-Williams approximation. This is the great advantage that the associate solution model is hardly to possess. If the associate solution model is expected to reach the limiting case of the ideal entropy of mixing, the Gibbs energy of formation of included complex species from their constituent components must approach positive infinity; it is physically unrealistic.

Assuming the MQMDPA is interpolated by the Kohler model, the pair interaction energies can be expressed as,

$$\Delta g_{AB}^{\langle 1 \rangle} = \Delta g_{AB}^{\langle 1 \rangle 0} + q_{AB}^{\langle 1 \rangle 10} \frac{X_{AA}}{X_{AA} + X_{BB} + X_{AB}^{\langle 1 \rangle} + X_{AB}^{\langle 2 \rangle}} \quad (48)$$

$$\Delta g_{AB}^{\langle 2 \rangle} = \Delta g_{AB}^{\langle 2 \rangle 0} + q_{AB}^{\langle 2 \rangle 01} \frac{X_{BB}}{X_{AA} + X_{BB} + X_{AB}^{\langle 1 \rangle} + X_{AB}^{\langle 2 \rangle}} \quad (49)$$

$$\Delta g_{AC}^{\langle 1 \rangle} = \Delta g_{AC}^{\langle 1 \rangle 0} + q_{AC}^{\langle 1 \rangle 10} \frac{X_{AA}}{X_{AA} + X_{CC} + X_{AC}^{\langle 1 \rangle} + X_{AC}^{\langle 2 \rangle}} \quad (50)$$

$$\Delta g_{AC}^{\langle 2 \rangle} = \Delta g_{AC}^{\langle 2 \rangle 0} + q_{AC}^{\langle 2 \rangle 01} \frac{X_{CC}}{X_{AA} + X_{CC} + X_{AC}^{\langle 1 \rangle} + X_{AC}^{\langle 2 \rangle}} \quad (51)$$

$$\Delta g_{BC}^{\langle 1 \rangle} \quad (52)$$

where $\Delta g_{AB}^{\langle 1 \rangle 0} = \Delta g_{AB}^{\langle 2 \rangle 0} = 2q_{AB}^{\langle 1 \rangle 10} = 2q_{AB}^{\langle 2 \rangle 01} = \Delta g_{AC}^{\langle 1 \rangle 0} = \Delta g_{AC}^{\langle 2 \rangle 0} = -2q_{AC}^{\langle 1 \rangle 10} = -2q_{AC}^{\langle 2 \rangle 01} = -30000 J/mol$ and $\Delta g_{BC}^{\langle 1 \rangle} = -6000 J/mol$ are assigned. These result in the entropy of mixing, enthalpy of mixing and Gibbs energy of mixing, as shown in Fig.3.

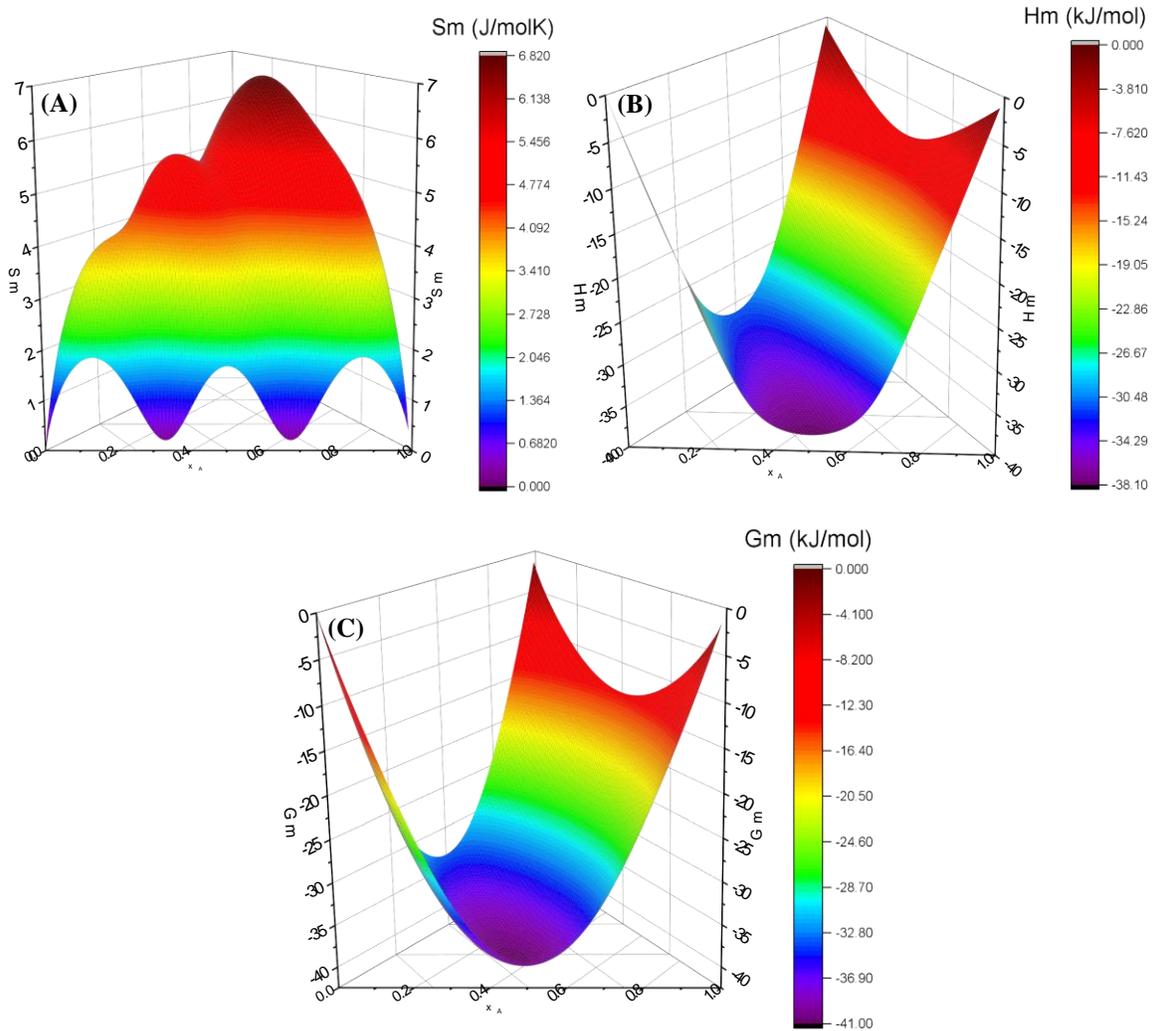

Fig.3 Thermodynamic properties of the A-B-C solution calculated over the whole composition space using the symmetrical Kohler model: (A) Configurational entropy; (B) Enthalpy of mixing; (C) Gibbs energy of mixing

If the MQMDPA is interpolated by the Kohler-Toop model with component A selected as the asymmetric, the interaction pair energies should thus be given as,

$$\Delta g_{AB}^{\langle 1 \rangle} = \Delta g_{AB}^{\langle 1 \rangle 0} + q_{AB}^{\langle 1 \rangle 10} X_{AA} \tag{53}$$

$$\Delta g_{AB}^{\langle 2 \rangle} = \Delta g_{AB}^{\langle 2 \rangle 0} + q_{AB}^{\langle 2 \rangle 01}(X_{BB} + X_{CC} + X_{BC}^{\langle 1 \rangle} + X_{BC}^{\langle 2 \rangle}) \tag{54}$$

$$\Delta g_{AC}^{\langle 1 \rangle} = \Delta g_{AC}^{\langle 1 \rangle 0} + q_{AC}^{\langle 1 \rangle 10} X_{AA} \tag{55}$$

$$\Delta g_{AC}^{\langle 2 \rangle} = \Delta g_{AC}^{\langle 2 \rangle 0} + q_{AC}^{\langle 2 \rangle 01}(X_{BB} + X_{CC} + X_{BC}^{\langle 1 \rangle} + X_{BC}^{\langle 2 \rangle}) \tag{56}$$

$$\Delta g_{BC}^{\langle 1 \rangle} \tag{57}$$

where all the coefficients are the same as those used in the above Kohler model. The calculated entropy of mixing, enthalpy of mixing and Gibbs energy of mixing are then given in Fig.4. It is seen from Fig.3 and Fig.4 that the calculated results are different in corresponding properties, in particular for the entropy of mixing.

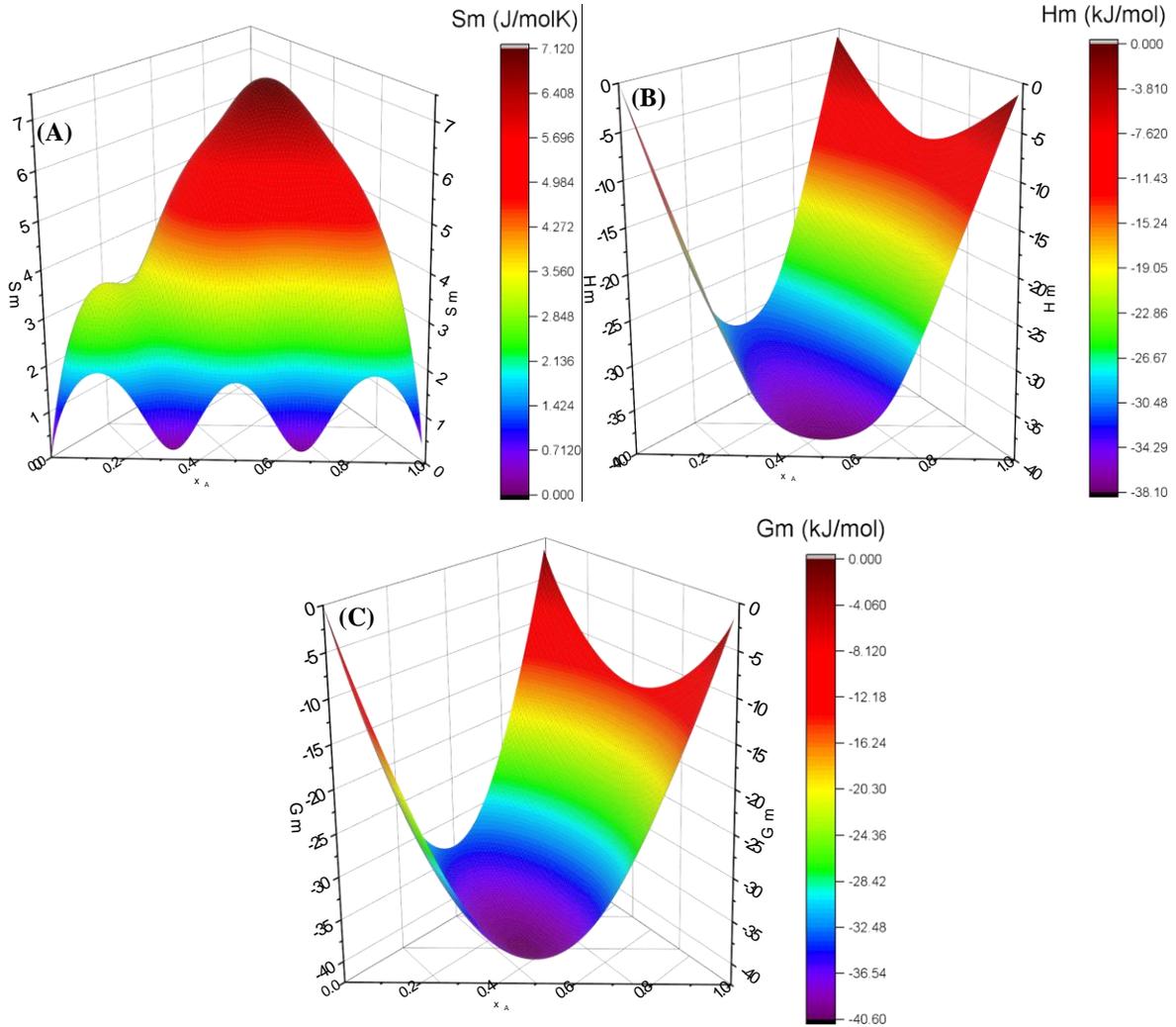

Fig.4 Thermodynamic properties of the A-B-C solution calculated over the whole composition space using the Kohler-Toop model with A as the asymmetric component: (A) Configurational entropy; (B) Enthalpy of mixing; (C) Gibbs energy of mixing

Once the MQMDPA is interpolated by the integration model, the pair interaction energies are supposed to have the following expressions,

$$\Delta g_{AB}^{\langle 1 \rangle} = \Delta g_{AB}^{\langle 1 \rangle 0} + q_{AB}^{\langle 1 \rangle 10}(X_{AA} + X_{CC}\xi_{AB}^2 + \xi_{AB}X_{AC}) \tag{58}$$

$$\Delta g_{AB}^{\langle 2 \rangle} = \Delta g_{AB}^{\langle 2 \rangle 0} + q_{AB}^{\langle 2 \rangle 01}(X_{BB} + X_{CC}(1-\xi_{AB})^2 + (1-\xi_{AB})(X_{BC}^{\langle 1 \rangle} + X_{BC}^{\langle 2 \rangle})) \tag{59}$$

$$\Delta g_{AC}^{\langle 1 \rangle} = \Delta g_{AC}^{\langle 1 \rangle 0} + q_{AC}^{\langle 1 \rangle 10}(X_{AA} + X_{BB}(1-\xi_{AC})^2 + (1-\xi_{AC})(X_{AB}^{\langle 1 \rangle} + X_{AB}^{\langle 2 \rangle})) \tag{60}$$

$$\Delta g_{AC}^{\langle 2 \rangle} = \Delta g_{AC}^{\langle 2 \rangle 0} + q_{AC}^{\langle 2 \rangle 01}(X_{CC} + X_{BB}\xi_{AC}^2 + (X_{BC}^{\langle 1 \rangle} + X_{BC}^{\langle 2 \rangle})\xi_{AC}) \tag{61}$$

$$\Delta g_{BC}^{\langle 1 \rangle} \tag{62}$$

where all the model parameters are the same as those used in the preceding interpolations except the similarity coefficients $\xi$. Based on equation (35), $\xi_{AB}$ and $\xi_{AC}$ are calculated as 0.0422 and 0.9341, respectively. $\xi_{BC}$ has the value of 0.6156, but is not used since the interaction energy of pair B-C is expressed as composition independent (or configuration independent). The calculated entropy of mixing, enthalpy of mixing and Gibbs energy of mixing are shown in Fig.5 for the ternary A-B-C solution. Compared to the preceding calculations, the present calculation is much closer to the one using the Kohler-Toop model.

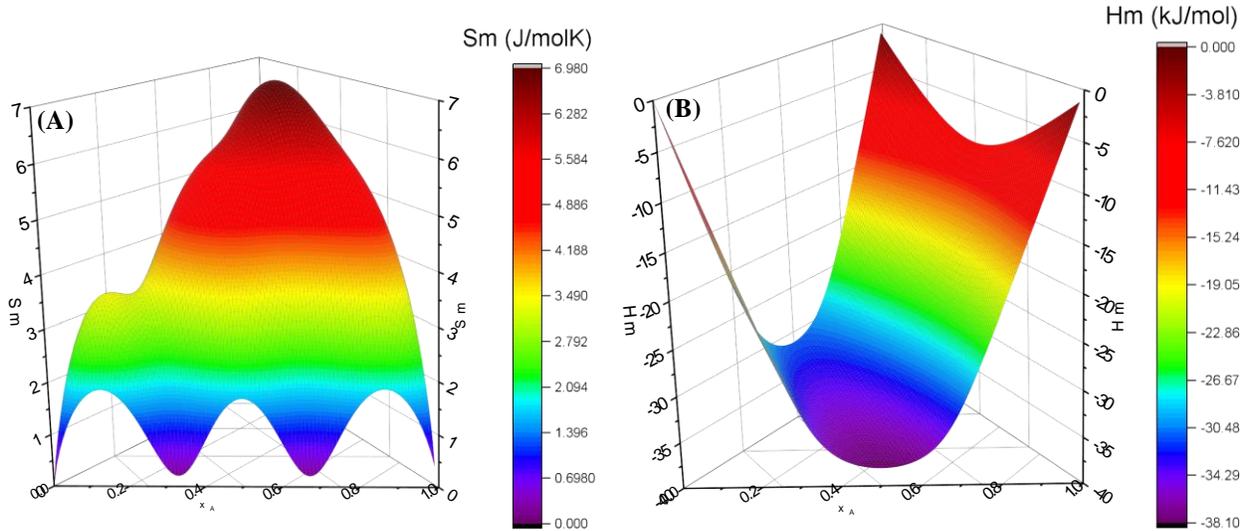

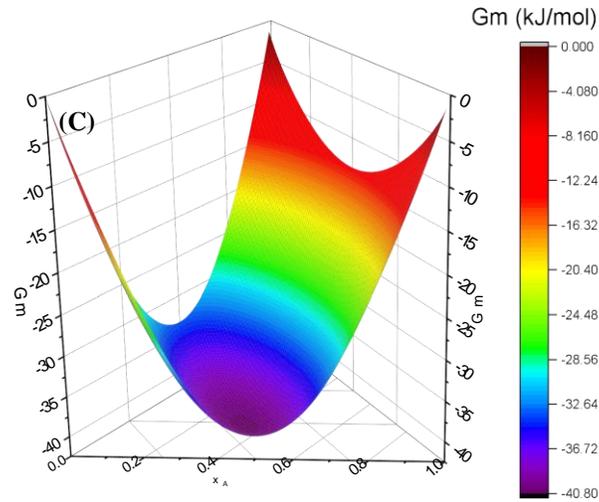

Fig.5 Thermodynamic properties of the A-B-C solution calculated over the whole composition space using the integration model: (A) Configurational entropy; (B) Enthalpy of mixing; (C) Gibbs energy of mixing

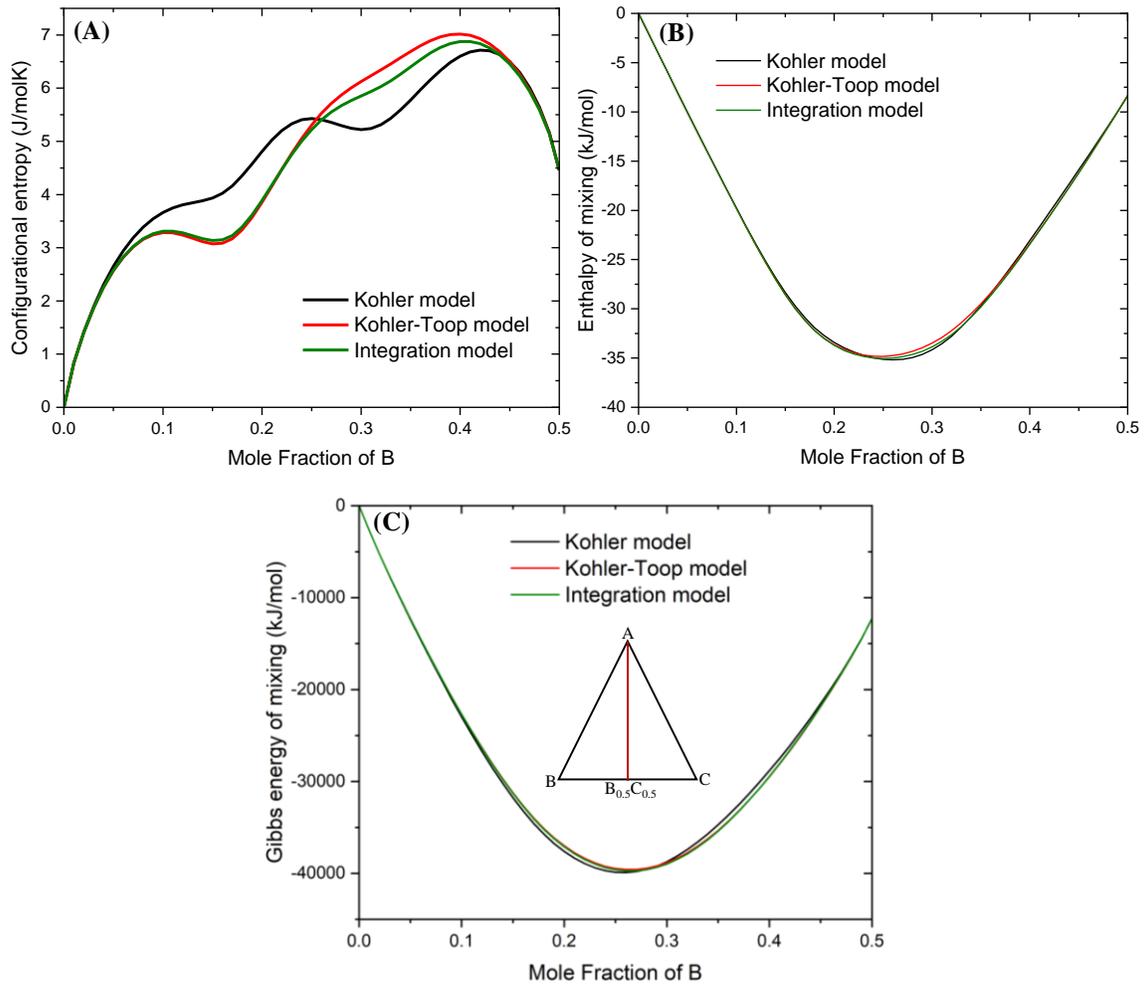

Fig.6 Thermodynamic properties of the A-B-C solution calculated along the A-$B_{0.5}C_{0.5}$ joints using the integration model: (A) Configurational entropy; (B) Enthalpy of mixing; (C) Gibbs energy of mixing

Much clearer comparisons among the calculations using the three interpolation models can be seen in Fig.6. These calculations are performed along the A-$B_{0.5}C_{0.5}$ joint. The calculated enthalpy of mixing and Gibbs energy of mixing seem to be very close in using the three interpolation models. However, there are obvious deviations among the calculated entropy of mixing (configurational entropy). The integration model has almost evolved into the asymmetric model driven by the system-dependent similarity coefficients. This is because there is similar energy level for the A-B and A-C pairs; they are very different from the B-C pair where much weaker interaction appears. As a result, the calculated entropy of mixing using the integration model approaches the one using the asymmetric Kohler-Toop model.

## 5. Concluding remarks

The modified quasichemcial model in the distinguishable-pair approximation, capable of treating solutions having manifold short-range orders, has been successfully extended for use in multicomponent systems in the present work. Three interpolation models, namely, Kohler, Toop and Chou, were initially introduced to express the pair interaction energies in ternary solutions by using the corresponding parameters optimized from their constituent binary solutions. A generic formalism for the combined Kohler-Toop model was later employed to allow complete freedom of choice to treat any ternary subsystem with a symmetric or asymmetric model. Meanwhile, a general Chou model was also used to combine energy expressions from all ternary subsystems into multicomponent systems. All of the interpolation methods have been applied to the pair interaction energies expressed either in terms of "coordination-equivalent" fractions or in terms of pair fractions. The combinatorial Kohler-Toop model requires human interference to select symmetric and asymmetric components according to some empirical rules, such as the chemical properties of components, the location of elements in a periodic table, or the valence of compounds. With all the analytical energy expressions, it is convenient to perform program implantation into current software. The Chou model introduces the similarity coefficient to characterize the deviation of properties from each two binary solutions with a co-member. The similarity coefficient provides the model to be dynamically shiftable among different interpolation models without any human interference. However, the similarity coefficient is an integral where the integrand includes internal variables and thus impossible to work out an analytical expression. This may impede the code implementation into current software with ready-made algorithm for multicomponent and multiphase phase equilibria calculations. An innovative algorithm may be required in order to possibly implant the Chou model in high efficiency. A subsequent paper may be published to describe how to extend the MQMDPA into two sublattices for use in reciprocal solutions.

## Acknowledgements

The Author Kun Wang would like to thank his wife and daughter for all their great supports and is also very grateful for being considerate of not spending much time with them.